\documentclass[letterpaper, 10 pt, conference]{ieeeconf}  

\IEEEoverridecommandlockouts                

\usepackage{graphicx,dblfloatfix} 
\usepackage{epsfig} 
\usepackage{mathptmx} 
\usepackage{times} 
\usepackage{amsmath} 
\usepackage{amssymb}  
\usepackage{cite}
 \usepackage{tikz}
 \usepackage{graphicx,color,psfrag}
\usetikzlibrary{shapes,arrows,shadows,calc,patterns,decorations.pathmorphing,decorations.markings}

\definecolor{tublue}{RGB}{0,166,214}
\definecolor{tuyellow}{RGB}{247,235,144}
\definecolor{blue}{RGB}{0,166,214}
 \definecolor{lblue}{RGB}{119,192,215}
 \definecolor{lred}{RGB}{236,127,44}

\title{\LARGE \bf
Active Power Control of Waked Wind Farms: Compensation of Turbine Saturation and Thrust Force Balance
}

\author{Jean Gonzalez Silva, Bart Doekemeijer, Riccardo Ferrari and Jan-Willem van Wingerden$^{*}$ 
\thanks{$^{*}$ Delft University of Technology, Delft, 2628CD The Netherlands
        {\tt\small \{J.GonzalezSilva, r.ferrari, B.M.Doekemeijer, J.W.vanWingerden\}@tudelft.nl} \newline
        {\small © 2021 \textit{This work has been accepted to ECC for publication under a Creative Commons Licence CC-BY-NC-ND}}}%
}

\begin{document}

\maketitle
\thispagestyle{empty}
\pagestyle{empty}

\begin{abstract}

Active power control regulates the total power generated by wind farms with the power consumed on the electricity grid. Due to wake effects, the available power is reduced and turbulence is increased at downstream wind turbines.
Such effects lead to a design challenge for wind farm control, where the delicate balance between supply and demand should be maintained, while considering the load balancing in the wind turbine structures.
We  propose  a control  architecture  based  on  simple  feedback  controllers  that adjusts the demanded power set points of individual wind turbines to compensate for turbine saturations and  to   balance   thrust  forces. For compensation purposes, the dynamics of power tracking in the wind turbines is approximated as a pure time-delay process, and the thrust force balance design is based on an identified linear model of the turbines.
In this paper, we show that the proposed control architecture
allows the generated power to track its reference even when turbines saturate, while the thrust forces are balanced. 
In addition, the result shows that the proposed power dispatch strategy, which considers thrust force balance, also avoids turbine saturation, being thus beneficial for energy production.
The effectiveness of the proposed feedback controller is demonstrated using high-fidelity computational fluid dynamics simulations of a small wind farm.

\end{abstract}

\section{INTRODUCTION}

As wind energy is an emerging renewable source in the world's energy portfolio, wind farm control systems play an ever more prominent role in the electricity grid.
Being installed in locations with more consistent wind, e.g. offshore areas, and having a higher generation capacity, the total power provided by wind farms has become more stable and reliable, which can track specific power demands. 
The integration of wind power with the electricity grid is an important, yet open challenge, in the current transition from fossil-fuel-based to renewable energy sources, in which several functions can still be improved and developed \cite{Veerseaau2027}.

In 2002, Rodriguez-Amenedo et al. \cite{amenedo2002} present an initial simulation study on active power control. 
A supervisory control system dispatches additional power demand signals to derated turbines in the situation where other turbines reach their maximum power production. 
Later publications such as Hansen et al. \cite{hansen2006}, Biegel et al. \cite{Biegel2013}, Ahmadyar, Verbi\v{c} \cite{ahmadyar2017} and Bay et al. \cite{bay2018} 
entail active power control algorithms. 
However, many algorithms in literature are tested in situations with sufficient power available in the wind, which is not a realistic assumption to make. 
Additionally, all of these algorithms were tested in low-fidelity simulation environments, and their applicability remains uncertain.


 As wind farm densification increases, turbulence and wake effects are clearly noticed \cite{fleming2013}. The first publication to perform a large-eddy simulation study on a waked wind farm was presented by Fleming et al. \cite{fleming2016}, where its power reference was equally dispatched among the turbines, although unequal amounts of power are available at each turbine due to wake formation. 
Since its publication, an increasing number of active power control algorithms have been tested in large-eddy simulations. Notable recent works are the model predictive controller of Boersma et al. \cite{boersma2019}, Shapiro et al. \cite{shapiro2017} and Van de Scheur and Boersma \cite{vandescheur2020}. The issue with all these controllers is their complexity, which is often a large barrier for adoption. Accordingly, Van Wingerden et al. \cite{vanWingerden2017} present a simple feedback controller that significantly improves the tracking behavior of the total power output of the farm in a large-eddy simulation. The controller compensates the wake effects (wind lulls) that may occur in one part of the farm, but saturation is not studied. 

Another challenge driven by the effects of the upstream turbines is related to the induced structural loads 
\cite{aho2012,knudsen2015}. In literature, there are many control algorithms that aim structural load mitigation. 
Optimal controllers are the most frequently adopted 
as in  Spudi\'c et al.\cite{Spudic2010}, Zhao et al. \cite{zhao2017} and Zhang et al. \cite{zhang2017}. 
 Also, loads are considered by Vali et. al \cite{vali2019}, as an extension of the compensation of power losses \cite{vanWingerden2017}. The power set points are distributed to minimize structural loads variations in waked wind farms based on
the measurements of the tower fore–aft bending moments. However, there exist scenarios, e.g. turbine saturation cases, where control goals of power tracking and load mitigation might conflict.



In this paper, we present a control architecture based on simple feedback controllers that adjusts turbine power set points to compensate turbine saturation, a simplification of the approach of [3], where pure time-delay process is considered. In addition, the control architecture balances thrust forces in a non-conflicting manner at only unsaturated turbines. The thrust force balance avoids the excessive lifetime reduction of specific turbines which can reduce the total maintenance cost \cite{zhang2017,Liu_2021ab}.


The main contribution of this paper is a simple feedback control architecture compared to the literature.
The architecture leads to an excellent wind farm power tracking performance and to thrust force balance evaluated in a high-waked scenario. We demonstrate 
that considering the thrust forces, not only structural loads are alleviated in an aggregate manner, but also the trade-off between demanded and available power is improved. 
Therefore, turbine saturation driven by the wake effects is avoided and handle at specific average wind speeds by the proposed architecture.

The structure of this paper is as follows. First, the wind turbine level is explored in  Section \ref{wtc}. Next, the proposed wind farm control architecture is presented in Section \ref{wfc}. In Section \ref{sim}, model identification is performed  and the simulation results of the proposed architecture are shown. The paper is concluded in Section \ref{conc}.

\section{WIND TURBINE LEVEL} \label{wtc}


\subsection{Power tracking with wind turbine controller}
 The turbine controller is synthesized to accurately track the demanded generator power of the $i$th single turbine $P^{dem}_i$, whenever possible. The pitch controller is set to exclusively regulate the generator speed to its rated value defined by the turbine machine. Due to the slow pitch dynamics, this setting is considered herein in order to have a fast control ability in terms of power tracking  \cite{kim2018}. Then, two operational modes can be distinguished based on the generator-torque controller.
\subsubsection{Control mode I - forced power reference tracking} \label{controlmod1}
The generator torque $\tau_{gen, \, tracking, \, i}$ necessary to meet a certain power demand, in the absence of time delays and relevant internal dynamics in the generator system, can be expressed as
\begin{equation} \label{torquetrack}
   \tau_{gen, \, tracking, \, i} = P^{dem}_i (\omega_{gen, \,i} \eta_{gen, \,i})^{-1},
\end{equation}
where $\omega_{gen, \,i}$ is the generator speed and $\eta_{gen, \,i}$ is the generator efficiency.
In the situation of a low $P^{dem}_i$, the rotor will spin up. 
Then, the generator speed is regulated by the pitch controller which the over spinning is avoided.
In contrast, the rotor will slow down with a high $P^{dem}_i$. 
This can be particularly problematic, since both the rotor speed and aerodynamic efficiency will continue to reduce and an increasingly higher generator torque is necessary to meet the desired power production. This behavior is unstable and leads to turbine shutdown. To prevent this, a secondary control mode is introduced.

\subsubsection{Control mode II - greedy control and power reference tracking}

The literature standard for wind turbine control is referred to as \emph{greedy control}. The generator torque control law that achieves the optimal tip-speed ratio $\lambda^{opt}$ can be defined as 
\begin{equation}
    \tau_{gen, \, greedy, \, i} = K  \omega_{gen, \, i} ^2,
\end{equation}
with the right value of $K$. This control law is stable and globally converges to the optimal power coefficient to maximize the turbine's power production for a large range of the tip speed ratio \cite{Johnson2006}. Additionally, regions and  linear transitions can be determined 
to guarantee stability and improve the control performance \cite{jonkman2009}; herein also considered as $\tau_{gen, \, greedy, \, i}$, for simplicity.  

Combining this with the power tracking control law yields
\begin{equation} \label{torquecomb}
    \tau_{gen, \, combined, \, i} = min (\tau_{gen, \, greedy, \, i}, \tau_{gen, \, tracking, \, i}).
\end{equation}

The generator-torque control law in Eq. \ref{torquecomb} ensures that in lower generator speed the turbine does not operate at a tip-speed ratio below $\lambda^{opt}$, addressing the instability issue described in subsection \ref{controlmod1}. In which, the turbine saturation is achieved when $P^{dem}_i > P^{greedy}_i$, where $P^{greedy}_i$ is the maximal power that can be produced in the defined condition. 

\subsection{Thrust force signals}

The thrust force $F_{T}$ expresses the main load generated by the wind into the turbine structure, and can be calculated by  
\begin{equation}
    F_{T} = \frac{1}{2} \rho \pi  R^2 v_{r}^2 C_{T}(\lambda, \, \theta),
\end{equation}
where $\rho$ is the air density, $R$ is the length of the blade, $v_r$ is the effective wind speed \cite{Liu_2021ac} on the rotor that can be estimated as in \cite{soltani2013}, $C_T$ is the thrust coefficient from pre-computed mapping,  $\theta$ is the collective pitch angle, and $\lambda$ is the tip speed ratio, defined by $\lambda = \frac{\omega_r R}{v_r}$, where $\omega_r$ is the rotor speed.


\begin{figure}[b!]
\centering
\includegraphics[width=1.02\linewidth]{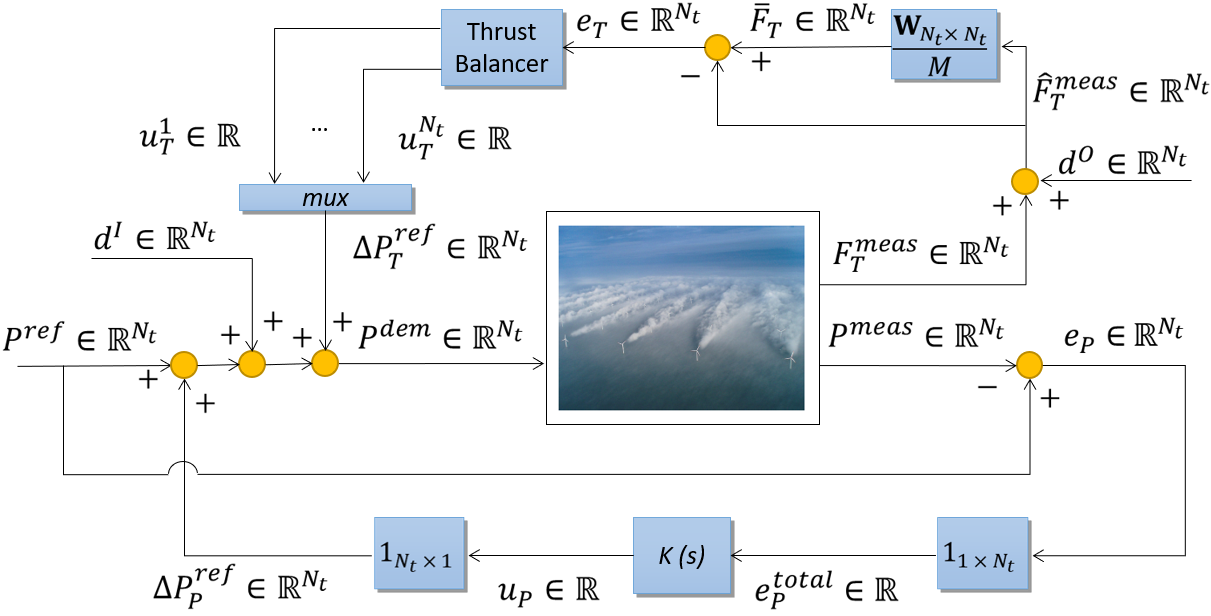}
\caption{Block diagram of the proposed wind farm control composed by a compensation control loop (CCL) at the bottom of the figure and  a thrust control loop (TCL) at the top of the figure (includes photograph of the Horns Rev 1 offshore wind farm, Christian Steiness).}
\label{BD}
\end{figure}

\section{WIND FARM LEVEL} \label{wfc}
The main ideas of our proposed controller are to maintain power tracking of the $N_t$ turbines in saturation scenarios and to balance the thrust forces to its mean value at the unsaturated turbines. Therefore, two loops are used: the compensation control loop (CCL) for turbine saturation; and the thrust control loop (TCL) composed by a thrust balancer.

As showing in the block diagram (Fig. \ref{BD}), the control signal $\Delta P^{ref}_P \in \mathbb{R}^{N_t}$ coming from the CCL automatically increase the set points of the individual turbines whenever turbine saturation occurs. On the other hand, $\Delta P^{ref}_T \in \mathbb{R}^{N_t}$ induces the balance of the measured thrust forces.

 For the CCL, the overall measured power $P^{meas}_{WF} \in \mathbb{R}$ is defined as $P^{meas}_{WF}=\mathbf{1}_{1 \times N_t} P^{meas} = \sum_{i=1}^{N_t} P^{meas}_{i}$, and the overall power reference $P^{ref}_{WF} \in \mathbb{R}$ is defined $P^{ref}_{WF}= \mathbf{1}_{1 \times N_t} P^{ref}  = \sum_{i=1}^{N_t} P^{ref}_{i}$, where $\mathbf{1}_{1 \times N_t}$ is defined as $[1 ... 1] \in \mathbb{R}^{1 \times N_t}$ and $\mathbf{1}_{N_t \times 1} = \mathbf{1}_{1 \times N_t}^T$. The vector $e_P \in \mathbb{R}^{N_t}$ contains the set point errors of the individual turbines, whereas $e^{total}_P = P^{ref}_{WF} - P^{meas}_{WF} \in \mathbb{R}$ is the overall tracking error of the wind farm.

For the TCL each of the measured thrust forces ${F}_{T, \, i}^{meas}\in \mathbb{R}$ is subtracted by the mean thrust force $\bar{F}_T \in \mathbb{R}$ to obtain the thrust force error vector $e_{T}\in \mathbb{R}^{N_t}$.  Then, we can formulate a LTI state space model in discrete time
as
\begin{equation} \label{ss1}
\begin{matrix}
F^{meas}_T(k+1)=\mathbf{A}F^{meas}_T (k) + \mathbf{B} (P^{ref}(k) + \Delta P_P^{ref} (k) \\ +\Delta P_T^{ref} (k) +  d^I (k) ),
\end{matrix}
\end{equation}
\begin{equation} \label{ss2}
e_T(k)=\left( \frac{1}{M} \mathbf{W}_{N_t \times N_t} - \mathbf{I} \right) (F^{meas}_T (k) +d^O (k)),
\end{equation}
where $k$ is the discrete time index of the simulation. $F^{meas}_T(k)$, $P^{ref}(k)$, $\Delta P_P^{ref}(k)$, and $\Delta P_T^{ref}(k)$ are $N_t \times 1$ vectors representing the measured thrust forces and the power distributions. The vectors $e_T(k)$, $d^I (k)$, and $d^O (k)$ are $N_t \times 1$ vectors representing the control thrust error, control disturbance, and measurement disturbance.
$\mathbf{I}$ is the $N_t \times N_t$ identity matrix.
The $N_t \times N_t$ matrices $\mathbf{A}$ and $\mathbf{B}$ contain model parameters is identified in next section for design a thrust balance controller via pole placement. Note that although the relationship with power set points and thrust forces is known to have a nonlinear behavior, a linear approximation around an operation region will be shown to be accurate enough for the present scope.

In addition, it was noticed that the TCL would unsuccessfully try to boost the generated power of saturated turbines. In the other hand, it would reduce the demanded power of the unsaturated turbines which might be used for compensation purposes. Consequently,   
 both loops might be competing with each other. Then, turbines that are saturated are removed from the thrust balancer, which $M$ is the amount of turbines that are not yet saturated, and
\begin{equation} \label{wm}
\mathbf{W}_{N_t \times N_t}= 
\begin{bmatrix} 
s_1 \, \, \, s_2 \, \, \, ... \, \, \, s_{N_t} \\ 
s_1 \, \, \, s_2 \, \, \, ... \, \, \,s_{N_t} \\
\vdots \\
s_1 \, \, \, s_2 \, \, \, ... \, \, \, s_{N_t}
\end{bmatrix}, \, \textrm{where } \begin{cases}
s_i=1 \textrm{, if turbine is} \\
\textrm{not saturated;} \\
s_i=0 \textrm{, if turbine is} \\
\textrm{saturated,}
\end{cases}
\end{equation}
is the balance weight matrix that removes the thrust forces of saturated turbines from the mean thrust force, and
\begin{equation}\label{zeroet}
e_{T,i}(k) = 0 \textrm{, if turbine is saturated.}
\end{equation}


The design of the two control loops are presented in the following subsections.

\subsection{Compensation of turbine saturation design}\label{pcd}

In the situation that $P^{dem}_i < P^{greedy}_i$ and above rated conditions, the input-output relationship is
$$
    P^{meas}_i(k) = \tau_{gen, \, tracking,\, i}(k) \omega_{gen,\, i}(k) \eta_{gen,\,i} 
    $$
\begin{equation}    
    = P^{dem}_i (k-1) (\omega_{gen,\, i}(k-1) \eta_{gen,\,i})^{-1} \omega_{gen,\, i}(k) \eta_{gen,\,i}.
\end{equation}
With a sufficiently high sampling rate, we can assume $\omega_{gen,\, i}(k) \approx \omega_{gen,\, i}(k-1)$ and therefore $ P^{meas}_i(k) \approx  P^{dem}_i(k-1)$. Thus, the wind turbine can be considered as pure delay systems with their time delay equal to the simulation sampling time $T_s$. 

The integral controller for the CCL with a positive feedback 
is simply defined as $ K(s) =\frac {K_I^{CCL}}{s}$,
where $K_I^{CCL}$ is the integrator gain.
In discrete time, it can be written as
\begin{equation}
    u_P(k) = u_P(k-1) + K_I^{CCL} e_P^{total}(k)T_s
\end{equation}
where $K_I^{CCL}=T_s^{-1}M^{-1}$ is by definition the optimal controller gain for pure time-delay systems. 
However, the optimal gain-scheduling, creates undesirable large variations on the demanded power in small wind farms due to the switch between very distinct gain values. Therefore, we propose the use of constant gain for small wind farms, as a fixed value $K_I^{CCL}=T_s^{-1}N_t^{-1}$, and let the integral action compensate for this decision. 
This controller achieves near-perfect tracking limited by the time delay of the measurements and rate limits imposed by the mechanics of the turbine machine - mainly the generator torque rate.
Integrator anti-windup is implemented when all turbines are saturated.

\subsection{Thrust balancer design} 

The integral controller can be elaborated as a dynamic partial state feedback. 
The system state is augmented by the integral error $e_I(k+1) = e_I(k) + e_T(k) T_s$ and the  feedback controller is defined as 
$\Delta P_T^{ref} = u_T(k) = \mathbf{K}_I^{TCL} e_I(k)$, where $\mathbf{K}_I^{TCL}$ is a diagonal matrix. 
In this way, the integral errors converge to zero but not the thrust forces. 
The closed loop system is 
\begin{equation}
\begin{matrix}
F^{meas}_T(k+1)=\mathbf{A}F^{meas}_T (k) + \mathbf{B}  \mathbf{K}_I^{TCL} e_I(k) + \mathbf{B} (P_{ref}(k) \\ + \Delta P_P^{ref}(k) +  d^I (k) ),
\end{matrix}
\end{equation}
\begin{equation}
e_I(k+1)= e_I(k) + \left( \frac{1}{M} \mathbf{W}_{N_t \times N_t} - \mathbf{I} \right) (F^{meas}_T (k) + d^O (k)) T_s .
\end{equation}


Then, it can be reorganized in a matrix formulation as
$$ \small{
    \begin{bmatrix}
    F^{meas}_T(k+1) \\
    e_I(k+1)
    \end{bmatrix}
    =
    \begin{bmatrix}
    \mathbf{A} & \mathbf{B}\mathbf{K}_I^{TCL} \\
    \left( \frac{1}{M} \mathbf{W}_{N_t \times N_t} - \mathbf{I} \right) T_s & \mathbf{I}
    \end{bmatrix}
   \begin{bmatrix}
    F^{meas}_T(k) \\
    e_I(k)
    \end{bmatrix}
    }
    $$
\begin{equation} +
     \begin{bmatrix}
    \mathbf{B}(P^{ref}(k) + \Delta P_P^{ref}(k) +  d^I (k) )\\
    (\frac{1}{M} \mathbf{W}_{N_t \times N_t} - \mathbf{I})T_s \, d_O (k)
    \end{bmatrix} .
\end{equation}
The closed loop characteristic equation is given by
\begin{equation}\label{characeq}
|\lambda \mathbf{I} - \mathbf{A}_{cl}| = 0   ,
\end{equation}
where
$$\mathbf{A}_{cl} = \begin{bmatrix}
    \mathbf{A} & \mathbf{B}\mathbf{K}_I^{TCL} \\
    \left( \frac{1}{M} \mathbf{W}_{N_t \times N_t} - \mathbf{I} \right) T_s & \mathbf{I}
    \end{bmatrix}.
   $$
   
The presence of a single uncontrollable pole on the unit circle will cause the obtained closed loop system to be only marginally stable, as expected. Indeed this property is structural and has been noted also in other works where load balancing is obtained by making individual loads track the average of all the loads \cite{Diao2014}.
However, gains can still be designed such as to stabilize the controllable poles. Also, once $det(\frac{1}{M} \mathbf{W}_{N_t \times N_t} - \mathbf{I})=0$ for all values of $\mathbf{W}_{N_t \times N_t}$, the controllable poles are still stable in switching $\mathbf{W}_{N_t \times N_t}$. 
As will be demonstrated by the simulations in the next section, the stability properties already guaranteed by the existing wind turbine controllers is maintained. 



\section{SIMULATIONS} \label{sim}

Large-eddy wind farm simulations were set on the high-fidelity Simulator fOr Wind Farm Applications (SOWFA) developed by the National Renewable Energy Laboratory (NREL) \cite{churchfield2012}.  For the sake of simplicity and fast simulations, the well known 5 MW reference wind turbine \cite{jonkman2009} is used along with the actuator disk model.  Through coupling SOWFA and MATLAB, using a network-based communication interface \cite{doekemeijer2019}, 
the identification was conducted for a proper control tuning in \ref{modeliden} and different scenarios were simulated in \ref{simul}. 

\subsection{Model Identification} \label{modeliden}
The identification process is conducted in open-loop by 
applying a step of 1 MW in the reference power of 2.5 MW with inflow wind speed of 12 m/s, so the measured power and thrust force responses are obtained.

\subsubsection{Power model}
The observed delay in the response of power is bigger than the adopted sampling time of $0.1$ s for the step of 1 MW applied in power reference. That is mainly due to the generator torque rate limit $15$ kNm/s imposed by the turbine. 
However, 
for low variations in the reference power, i.e lower than $166,667$ W / $0.1$ s, the measured power can be considered as a near-perfect delay of one sample time from the power reference. Therefore, the turbine can be considered as a pure time-delay system and the power control design in Subsection \ref{pcd} can be followed.


\subsubsection{Thrust force model}

A first-order model
\begin{equation}\label{firstom}
  \frac{F^{meas}_{T, i} (s)}{ P^{dem}_{i} (s)} = \frac{K_1}{T_1 s + 1}
\end{equation}
is identified with 84\% of fit, after a new tuning of the PID pitch controller ($K_D=0$, $K_P=1.82620057$ and $K_I=0.19566438$) to reduce the observed oscillations.


Applying the forward Euler discretization at (\ref{firstom}) yields 
\begin{equation}
  F^{meas}_{T, i} (k+1) = \left(1-\frac{T_s}{T_1}\right)F^{meas}_{T, i} (k) + \frac{T_s \, K_1}{T_1} P^{dem}_{i} (k).
\end{equation}

Finally, the matrices $\mathbf{A}$ and $\mathbf{B}$ can be written as 
$$ \mathbf{A} = diag\left(\left(1-\frac{T_s}{T_1}\right), \, ... \, , \, \left(1-\frac{T_s}{T_1}\right)\right)$$ 
$$ \mathbf{B} = diag\left(\frac{T_s \, K_1}{T_1}, \, ...\, , \, \frac{T_s \, K_1}{T_1} \right).$$

In this way, we model the thrust forces in a decentralized manner. Although changes in the demanded power from the upstream turbines influences the thrust forces of downstream turbines, the designed architecture with integral action, assuming the knowledge of all the thrust forces, eliminates the steady state error. Based on the identified model, the controllable poles of the characteristic equation of the close-loop (Eq. \ref{characeq}) can be placed. In this work $\mathbf{K}_I^{TCL} = 0.5 \mathbf{I}$ stabilizes the controllable poles with an overdamped behavior.

\subsection{Power tracking and thrust force balance}\label{simul}
\begin{figure}[b!]
\centering
\begin{center}
\includegraphics[width=\linewidth]{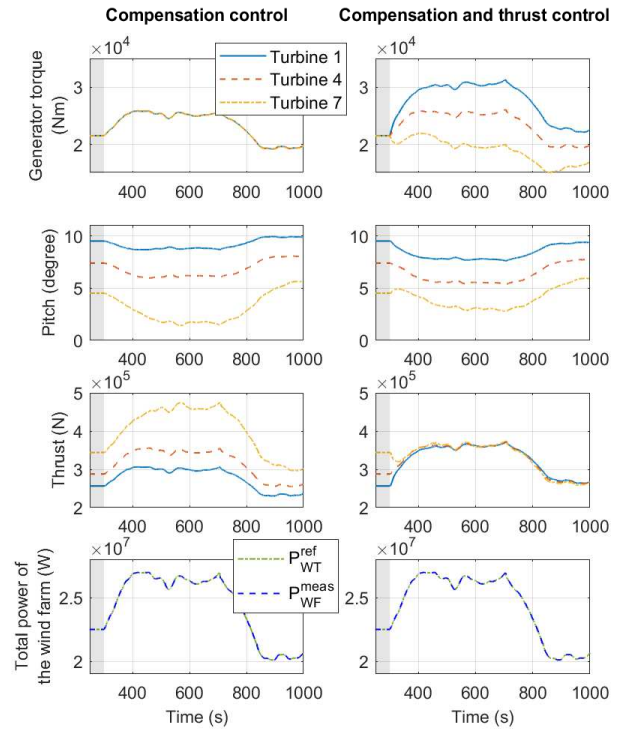}
\end{center}
\caption{Generator torque, pitch angles, thrust forces and total power of the wind farm with the compensation and thrust control loop at average wind speed of 13 m/s. The RMS error in power tracking are 3524.9 W and 3533.7 W, respectively.}
\label{varTCL}
\end{figure}
\begin{figure*}[t!]
\centering
\includegraphics[width=0.85\linewidth]{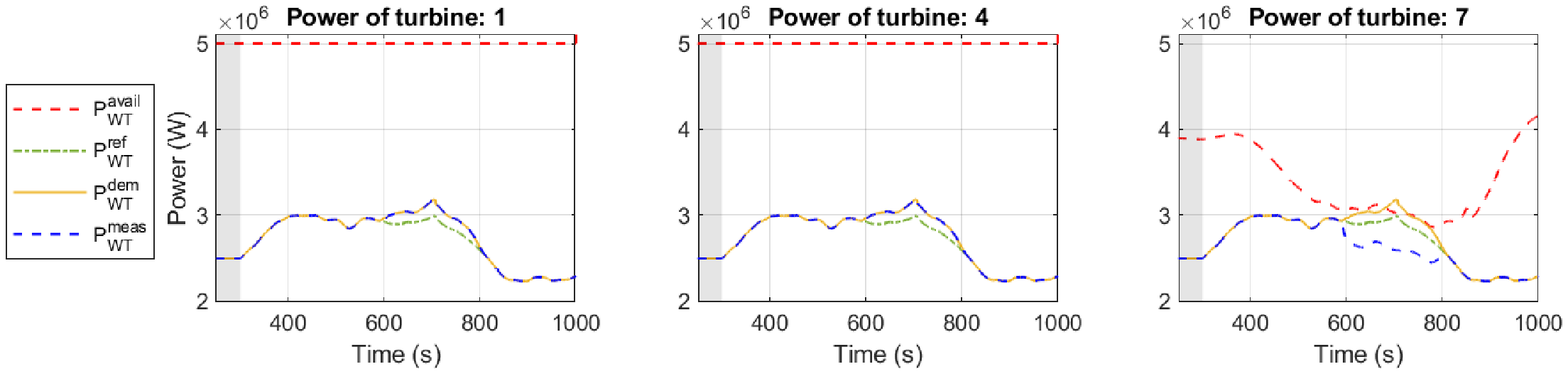}
\label{CCL}
\end{figure*}
\begin{figure*}[t!]
\centering
\includegraphics[width=0.85\linewidth]{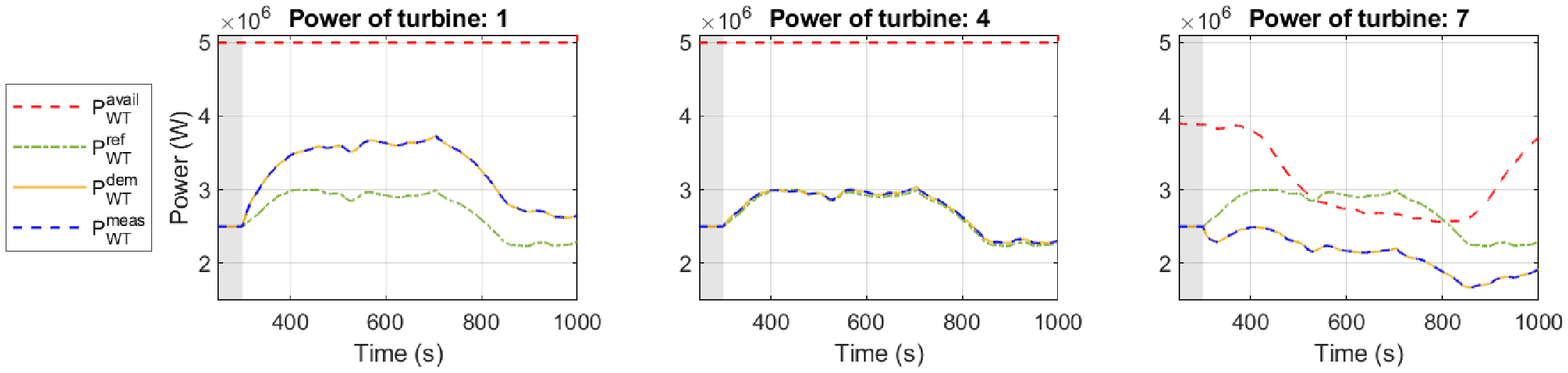}
\caption{Power of the individual wind turbines at average wind speed of 12.75 m/s with setting case 1 in the upper plots and with setting case 2 in the down plots. Note that the setting case 2 prevents saturation to occur.}
\label{TCL}
\end{figure*}

\begin{figure*}[t!]
\centering
\includegraphics[width=0.85\linewidth]{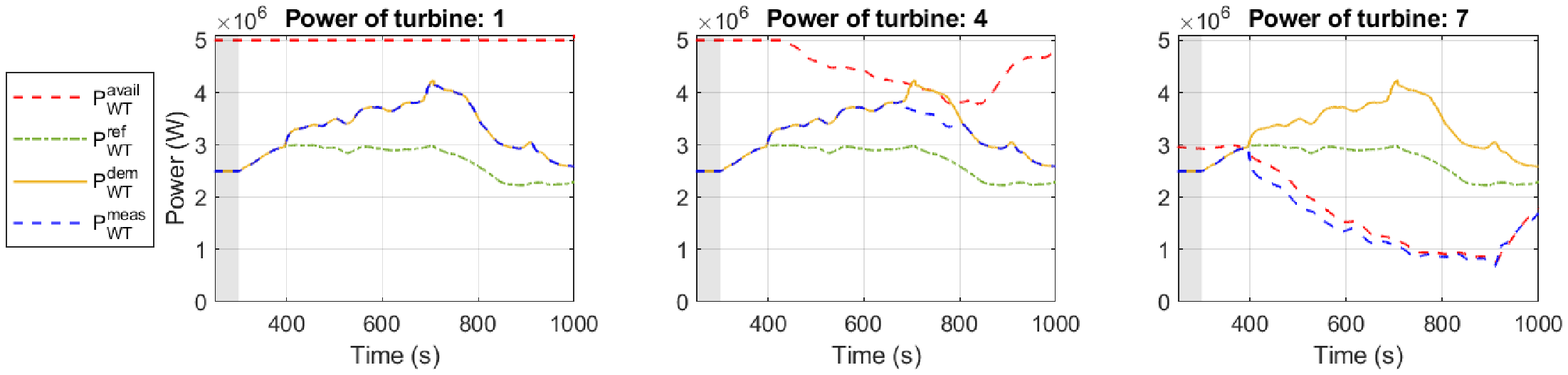}
\label{CCLl}
\end{figure*}

\begin{figure*}[t!]
\centering
\includegraphics[width=0.85\linewidth]{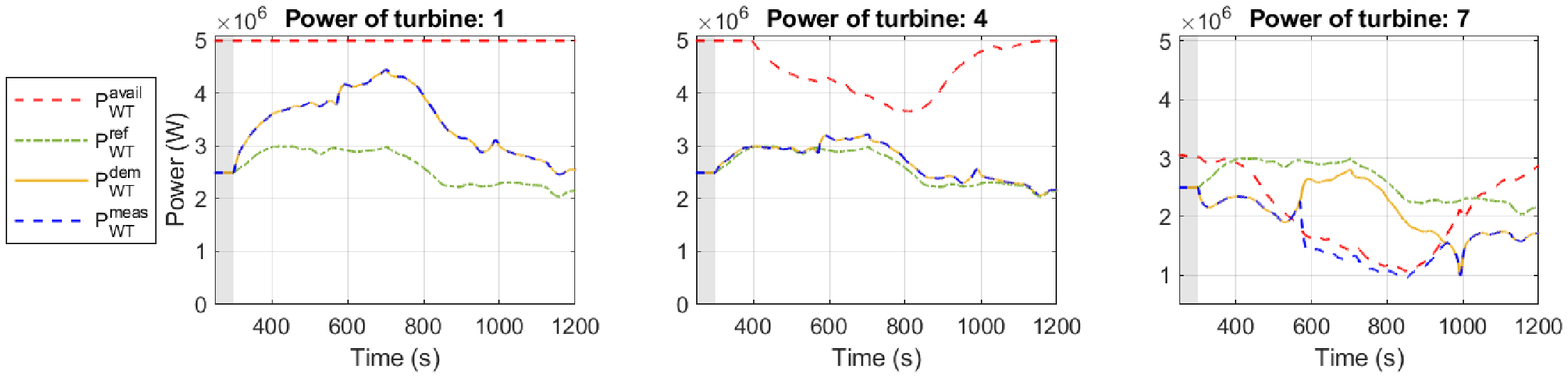}
\caption{Power of the individual wind turbines at average wind speed of 12.25 m/s with setting case 1 in the upper plots and with setting case 2 in the down plots. Note that turbine saturation occurs in both setting cases.}
\label{TCLl}
\end{figure*}

A small farm composed of 3 rows and 3 columns of turbines, set at a distance of 5 rotor diameters from each other, is simulated. No turbulence intensity is imposed on the wind inflow in order to maximize the wake effects. 
Initially, the de-rating case at $50\%$ of the rated power is considered and the distribution of the overall power reference is uniform among the turbines. 
Then, a time-varying power reference based on the normalized standard test signal \cite{pilong2013} and the wind farm controllers are considered after 300s. The demanded power from the TSO is set to reach the maximum of $70\%$ of the total rated power. 
The available power herein is extrapolated using the maximum power coefficient $C_{p_{max}}$ and illustrated by $P_{i}^{avail}(v_r)=min(0.5 \rho \pi  R^2 v_r^3 C_{p_{max}}, P_{i}^{rated}),$
where the effective wind speed $v_r$ is estimated by the I\&I technique \cite{ortega2011}.
To assess the performance of the proposed controller, the setting case 1 with only the CCL on and the setting case 2 with both the CCL and TCL on are simulated at different average inflow wind speeds. The result is illustrated by plotting only the first column of turbines for the sake of brevity, the other columns behaving similarly.

First, at average inflow wind speed of 13 m/s to 14 m/s, even though the wake is strong, turbine saturation does not occur in neither of the setting cases. The thrust forces converges to their mean value while power is tracked in the setting case 2, as shown in Fig. \ref{varTCL}. This would have a significant impact on the alleviation of the aggregated structural load, which will prolong the lifespan of specific wind turbines.


Then, at the average inflow wind speed of 12.75 m/s, in the setting case 1, the third row of turbines get saturated, and the power losses are compensated. On the other side, no turbine saturation is obtained in the setting case 2, where the turbine saturation was avoided by balancing the thrust force, as presented in Fig. \ref{TCL}. 

\begin{figure}[h!]
\centering
\includegraphics[width=\linewidth]{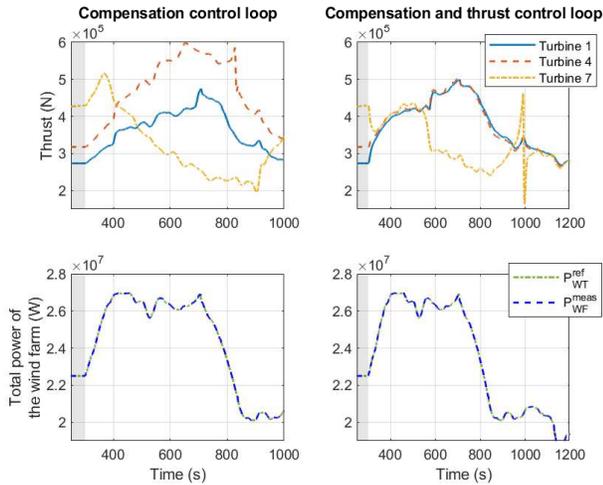}
\caption{Thrust forces with the compensation and thrust control loop at average wind speed of 12.25 m/s. The RMS error in power tracking are 6678.9 W and 5197.7 W, respectively, in the period of 300 s to 1000 s}
\label{T10}
\end{figure}

Finally, at the average inflow wind speed of 12.25 m/s, turbine saturation is found in both setting cases.
In the setting case 1, the boost of the turbine reference power also leads to a saturation of the turbines in the second row, seen at Fig. \ref{TCLl}.
In the setting case 2, even though the thrust control loop influences the trade-off between demanded and available power in individual turbines, turbines from the third row saturate. Then the compensation control loop is activated while the thrust control loop still balances the thrust forces in the first two rows until the third becomes unsaturated again - Fig. \ref{T10}. At that moment about $950 s$, because of the high rate of available power and the rotor inertia, the thrust in third row still increase even though the thrust balancer is acting, and a sudden drop is observed about $1000 s$ due to the strong pitch actuation from the wind turbine controller. Then, once the turbine reaches above rated condition, all the thrust forces are balanced.


\section{CONCLUSIONS} \label{conc}

We have shown that the proposed simple architecture for active power control compensates turbine saturation and balances the thrust forces at the unsaturated turbines in a non-conflicted way. The architecture alleviates the loads in an aggregated manner, and demonstrates a cooperative pattern of dispatch power, avoiding turbine saturation in wake scenarios. 

In the future we are going to further investigate prevention strategies at wind turbine level for transitions of operation regions to avoid undesirable jumps in thrust as observed and the cooperative pattern. Moreover, we are going to consider the prognosis of the remaining useful life and fault case scenarios in the dispatch power problem. 
\nocite{sorensen2005}





\section*{ACKNOWLEDGMENT}

The authors would like to acknowledge the WATEREYE project (grant no. 851207). This project has received funding from the European Union Horizon 2020 research and innovation programme under the call H2020-LC-SC3-2019-RES-TwoStages.


\bibliographystyle{IEEEtran}
\bibliography{IEEEabrv,IEEEexample}

\end{document}